\documentclass[10pt,aps,twocolumn,prd,superscriptaddress,showpacs,nofootinbib,noshowkeys,floatfix,preprintnumbers]{revtex4}
\usepackage{graphics,graphicx}
\usepackage{amsmath, amssymb}
\usepackage{multirow}
\usepackage{longtable}
\usepackage{color}

\usepackage[normalem]{ulem}  

\begin{document}

\title{
Strangeness fluctuations from $K-\pi$ interactions}


\date{\today}
\author{Bengt Friman}
\affiliation{GSI, Helmholzzentrum f\"{u}r Schwerionenforschung,
Planckstr. 1, D-64291 Darmstadt, Germany}
\author{Pok Man Lo}
\affiliation{Institute of Theoretical Physics, University of Wroclaw,
PL-50204 Wroc\l aw, Poland}
\author{Micha\l {} Marczenko}
\affiliation{Institute of Theoretical Physics, University of Wroclaw,
PL-50204 Wroc\l aw, Poland}
\author{Krzysztof Redlich}
\affiliation{Institute of Theoretical Physics, University of Wroclaw,
PL-50204 Wroc\l aw, Poland}
\affiliation{Extreme Matter Institute EMMI, GSI,
Planckstr. 1, D-64291 Darmstadt, Germany}
\author{Chihiro Sasaki}
\affiliation{Institute of Theoretical Physics, University of Wroclaw,
PL-50204 Wroc\l aw, Poland}
\affiliation{Frankfurt Institute for Advanced Studies, D-60438 Frankfurt am Main, Germany}

\begin{abstract}

Motivated by recent lattice QCD studies, we explore the effects of interactions on strangeness fluctuations in strongly interacting matter at finite temperature. We focus on S-wave $K\pi$ scattering and discuss the role of the $K_0^*(800)$ and $K^*(1430)$ resonances within the S-matrix formulation of thermodynamics. Using the empirical $K\pi$ phase shifts as input, we find that the $K\pi$ S-wave interactions provide part of the missing contribution to the strangeness susceptibility. Moreover, it is shown that the simplified treatment of the interactions in this channel, employed in the hadron resonance gas approach, leads to a systematic overestimate of the strangeness fluctuations.

\end{abstract}


\pacs{24.10.Pa, 25.70.Bc, 25.70.Ef, 25.75.-q}

\maketitle 

\section{missing interaction strength in strange sector}

A recent study, comparing QCD thermodynamics obtained on the lattice with the hadron resonance gas (HRG) model~\cite{Bazavov:2014xya}, indicates that additional interaction strength, beyond that embodied by well established strange resonances~\cite{Agashe:2014kda}, may be needed in the HRG model to remove disparities with the lattice results. In particular, the HRG results for the strangeness  and mixed strangeness-baryon number susceptibilities ($\chi_{SS}$ and $\chi_{BS}$) are clearly below those of the lattice, while the results for the thermodynamic pressure and the baryon number susceptibility are in good agreement. 

This motivates the search for hitherto unknown strange hadrons, which could reduce or eliminate this discrepancy. In the PDG database, there are around twenty unconfirmed states with a mass below $2.0 \, {\rm GeV}$. Although these are not established resonances, the interactions in the corresponding scattering channels may yield important contribution to thermodynamic quantities.

\begin{figure*}[t]
 \includegraphics[width=0.49\textwidth]{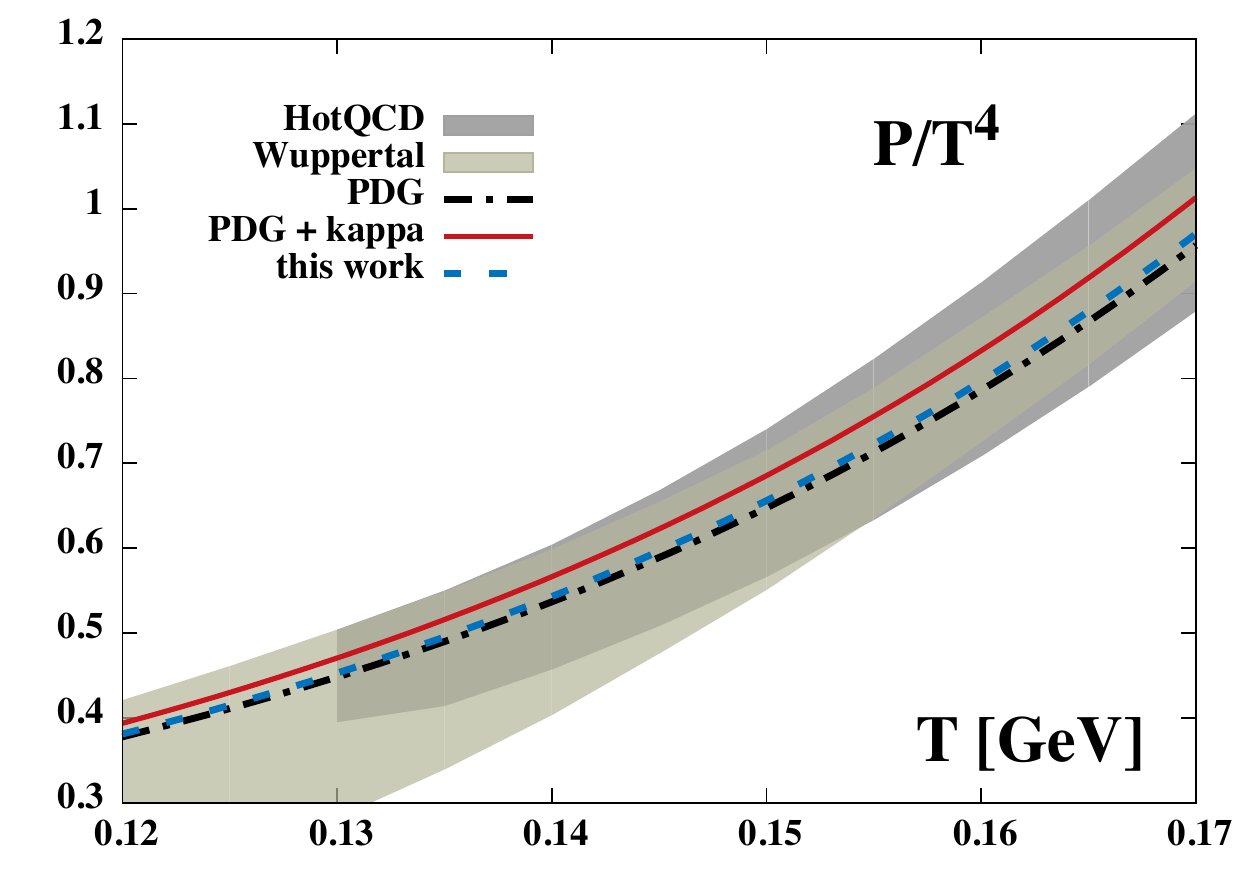}
 \includegraphics[width=0.49\textwidth]{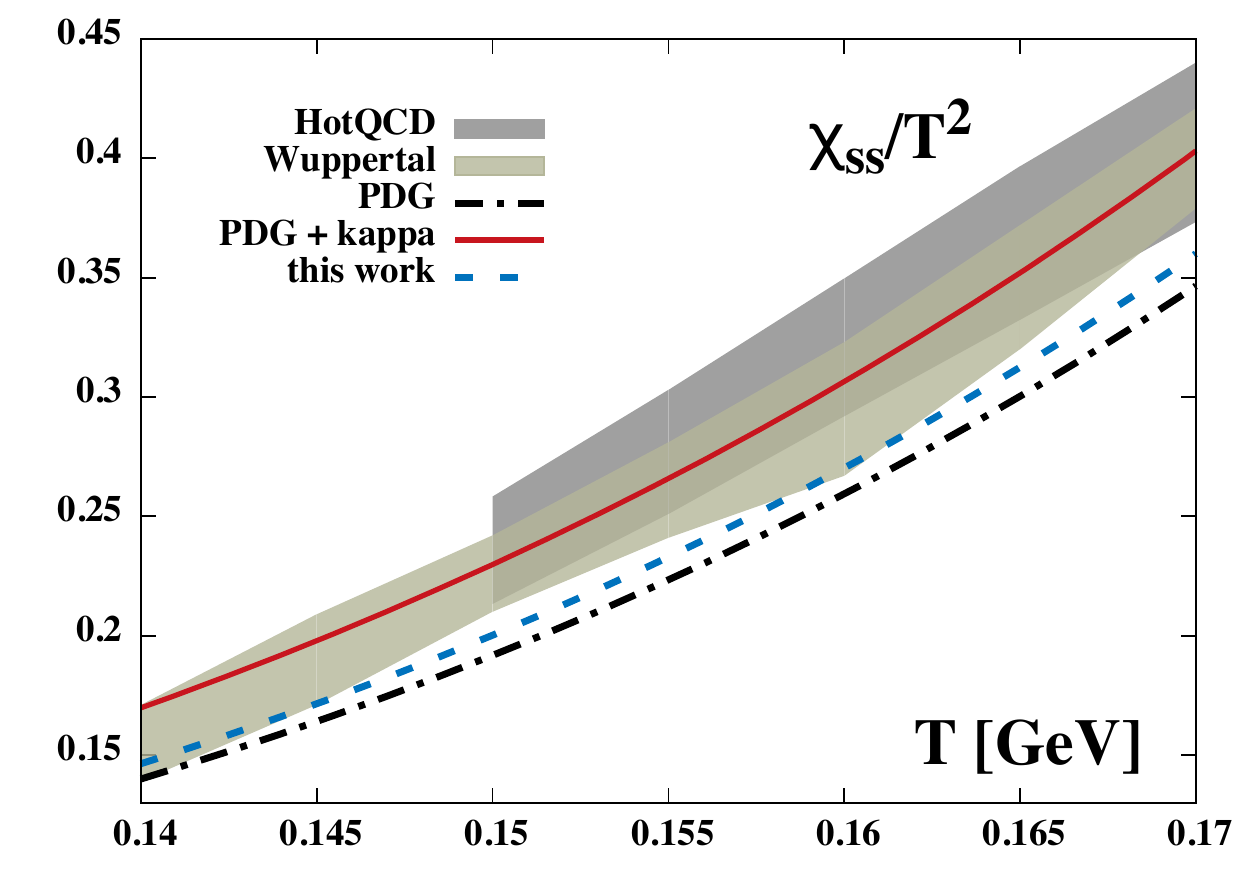}
 \caption{(color online). Left: The thermodynamic pressure (normalized to $T^4$) computed in HRG using only established resonances (PDG, broken dashed line) and adding the unconfirmed $\kappa$ (PDG + kappa, full line). The lattice results on pressure are from Refs. \cite{Bazavov:2014pvz, Borsanyi:2013bia}. The dashed line shows the results where the $\kappa$ channel is treated within the S-matrix formulation (see text). Right: The corresponding results for strange susceptibility $\chi_{SS}$ (normalized to $T^2$). The lattice results on fluctuations are from Ref. \cite{Bazavov:2012jq,Borsanyi:2011sw}. }
 \label{fig:hrg}
\end{figure*}

More generally, a possible origin of the discrepancy is interaction strength in channels carrying net strangeness that so far have not been accounted for. Given the corresponding empirical scattering phase shifts, both confirmed and unconfirmed resonances as well as non-resonant interactions can be handled in a unified, model-independent way, using the S-matrix approach of Ref.~\cite{Dashen:1969ep}. 

The strange scalar channel, with the unconfirmed $K^*_0(800)$ resonance, a.k.a. $\kappa$, is a prime candidate. Since the corresponding phase shifts for S-wave $K\pi$ scattering are fairly well determined, this channel is well suited for the S-matrix approach. In addition, the counterpart of $\kappa$ in the scalar-isoscalar channel, the $f_0(500)$, a.k.a. $\sigma$, though considered to be established \cite{Pelaez:2013jp}, is unlike a typical resonance. Since the $\pi\pi$ S-wave phase shifts are known with reasonable accuracy, also this channel is a prime candidate for the S-matrix approach to thermodynamics. In this study we focus on the strange scalar channel and its contribution to strangeness susceptibilities. 

With the relatively low mass of the interaction strength in the $\kappa$ channel, it potentially has a large impact on the thermodynamics, in particular on $\chi_{SS}$, owing to the moderate suppression by the Boltzmann factor. In Fig.~\ref{fig:hrg}, we illustrate the effect of the $\kappa$ resonance on pressure and strangeness fluctuation within the HRG approach. For the PDG particle spectrum, we use only confirmed baryons (i.e. three and four star resonances) and established mesons. The contribution of $\kappa$ to the thermodynamics is approximated by that of an ideal gas of zero-width mesons with mass $m_\kappa = 0.682 \, {\rm GeV}$ and degeneracy four. Indeed, the inclusion of this single state improves the HRG result on $\chi_{SS}$ dramatically, while the agreement in the thermodynamic pressure persists. However, owing to the fairly large width, the treatment of the $\kappa$ resonance as a zero-width particle is questionable. Consequently, a systematic approach, where all interaction effects are treated consistently, is called for.

In this paper, we assess the effect of interactions in the $\kappa$ channel on the thermodynamics using the S-matrix approach~\cite{Dashen:1969ep}. For elastic scattering, the resulting expression reduces to the Beth-Uhlenbeck form for the second virial coefficient, expressed in terms of the scattering phase shift~\cite{Beth:1937zz}. In Ref. \cite{Weinhold:1997ig} this scheme was applied to compute the contribution of $\pi N$ interactions to the baryon number and the $\pi$ transverse momentum spectrum in hadronic matter at moderate temperatures and densities. The method yields an effective spectral weight, which is relevant for the partition sum and thus allows one to compute the interaction contribution to various thermodynamic observables.

The paper is organized as follows. In section II we  describe the parametrization of the empirical $K \pi$ phase shifts. In particular, we implement the constraint provided by the empirical scattering length in this channel. Moreover, we discuss the connection between the phase shifts and thermodynamic quantities. In section III we apply this formalism to study the effect of interactions in the $\kappa$ channel. We compare these results with those of the standard HRG approach, where the $\kappa$ meson is treated as a Breit-Wigner resonance with an energy-independent width and study the influence of width and scattering length on strangeness fluctuations. In the final section, we present our conclusions.

\section{S-wave scattering and the S-matrix approach}

The scattering phase shift contains the necessary physical information to study resonances using scattering data. Since the $\kappa$ meson has the quantum numbers $I(J^P)=\frac{1}{2}(0^+)$, the relevant phase shift to consider is that of kaon-pion scattering in the S-wave, isospin $I=1/2$ channel (${\delta_0^{{1}/{2}}}$). We shall begin by collecting some basic field theoretical results to establish the connection between resonance width and the scattering phase shift pertinent to the study of the $\kappa$ resonance.

\subsection{S-wave decay of resonance}

Consider the decay of a scalar particle $\Phi \rightarrow \phi + \phi$, through an interaction term $\mathcal{L}_I = - g \, \Phi \phi^2 $. The self-energy of $\Phi$ is, to leading order in perturbation theory, given by

\begin{align}
	\Sigma_\Phi(k^2) &= 2 i g^2 \int \frac{d^4 l}{(2 \pi)^4} \frac{1}{l^2-m^2} \frac{1}{(l-k)^2-m^2} \nonumber\\
			 &=\frac{1}{(2 \pi)^4} \, 2 g^2 \pi^2 \int_0^1 dx \, \ln[(m^2-x(1-x) \, k^2) \, \pi],
\end{align}

\noindent where $m$ is the mass of the particle $\phi$. The decay rate is obtained from the imaginary part of $\Sigma_\Phi(k^2)$. It is clear from the above expression that the self-energy will develop an imaginary part when the invariant four momentum exceeds the threshold, i.e. $s = k^2 > 4 m^2$. The width of $\Phi$ in this model reads

\begin{align}
	\gamma(s) &= \frac{-\Sigma^I_\Phi(s)}{\sqrt{s}} \nonumber \\
				       &= \frac{g^2}{8 \pi} \, \theta[s- (2 m)^2] \, \sqrt{1-(2 m)^2/s} \, \frac{1}{\sqrt{s}}.
\end{align}

In the more general case of S-wave decay of a resonance into two particles with different masses $m_1$ and $m_2$, one finds

\begin{align}
	\label{eqn:width_para}
	\gamma(s) &= \frac{\alpha}{2} \, \theta[s- m_{th}^2] \, \frac{P_{CM}(s)}{s} \nonumber \\
	P_{CM}(s) &= \frac{1}{2} \, \sqrt{s} \, \sqrt{1-m_{th}^2/s} \, \sqrt{1-\Delta m^2/s} \ \\
	\Delta m &= m_1 - m_2  \nonumber \\ 
	m_{th} &= m_1 + m_2. \nonumber
\end{align}

\noindent Here we have introduced the notations $\alpha = g^2/4 \pi$ and the center of mass momentum $P_{CM}(s)$. Note the different symmetry factor in Eq.~\eqref{eqn:width_para}, owing to the distinguishable particles in the final state. As we show below, the energy dependence of the width $\gamma(s)$ is crucial for reproducing the S-wave phase shifts near threshold. 

\subsection{Parametrization of the S-wave $K$$\pi$ phase shift}

Although the expression for the decay width in Eq.~\eqref{eqn:width_para} is obtained from a perturbative one-loop calculation, it provides a general form for parametrizing the phase shifts of S-wave scattering. We account for the contribution of the two lightest $0^+$ strange resonances, $\kappa$ and $K^*_0(1430)$, to the $K$$\pi$ phase shifts by using an energy-dependent Breit-Wigner form 

\begin{align}
	\label{eqn:ps_para1}
	\delta_{\rm res}(s) &= \tan^{-1}(\frac{-\sqrt{s}\, \gamma(s)}{s-M_0^2}) \nonumber \\
	\gamma(s) &= \frac{\alpha}{2} \, \theta[s- m_{th}^2] \, \frac{P_{CM}(s)}{s}
\end{align}

\noindent for each resonance. Here $\alpha$ and $M_0$ are free parameters, which are fitted to the data.

However, in addition to the resonance contribution, a repulsive background contribution is needed for a successful description of the empirical phase shifts in this channel. Following \cite{Ishida:1997wn}, we parametrize the background using the phase shift of a hard sphere 

\begin{align}
	\label{eqn:ps_para2}
	\delta_{BG}(s) = - r_c P_{CM}(s),
\end{align}

\noindent where $r_c$ is the radius of the repulsive core. The total phase shift ${\delta_0^{{1}/{2}}}$ is given by the sum of the resonance contributions and the background

\begin{align}
	\label{eqn:ps_para3}
	{\delta_0^{{1}/{2}}} = \delta_{\kappa} + \delta_{K^*_0} + \delta_{BG}.
\end{align}

\begin{figure}
\centering
 \includegraphics[width= 0.49\textwidth]{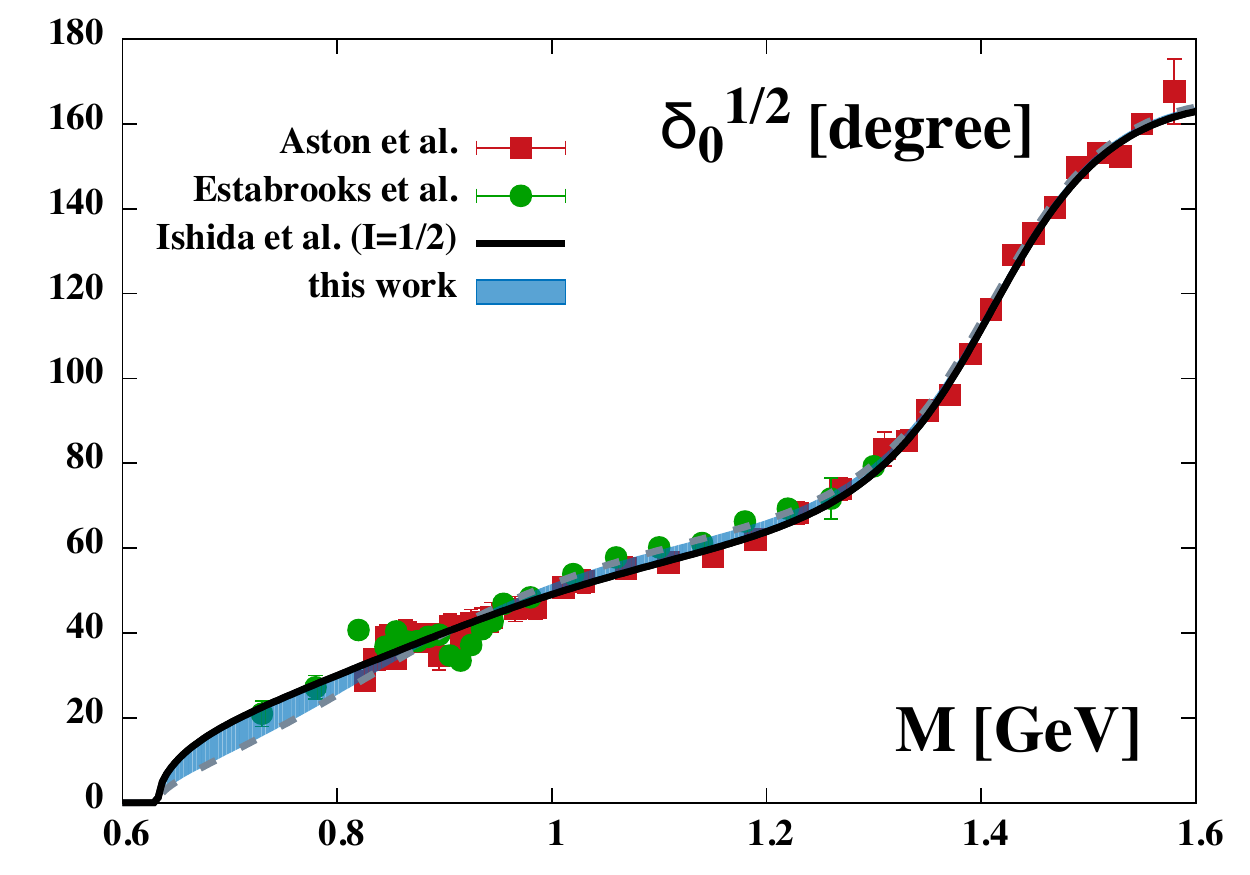}
 \includegraphics[width= 0.49\textwidth]{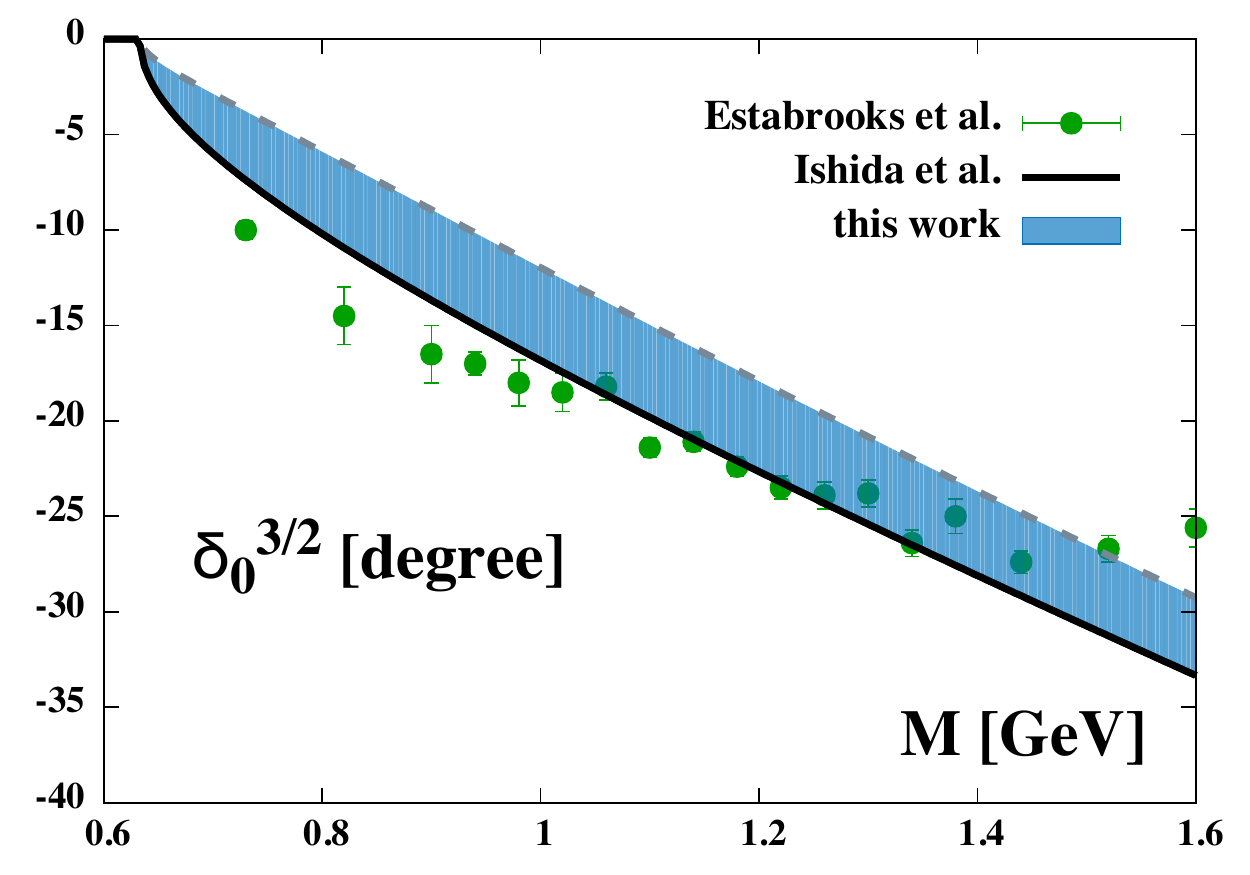}
 \caption{(color online). Top: The $K$$\pi$ scattering phase shift in the S-wave, isospin $I=1/2$ channel. The experimental results are obtained from Ref. \cite{ Aston:1987ir, Estabrooks:1977xe}. The solid line corresponds to the parametrization discussed in Eqs.~\eqref{eqn:ps_para1}~-~\eqref{eqn:ps_para3} using the parameters depicted in Table~\ref{tab:table1}. The band corresponds to different values of the scattering length, obtained by adjusting the regulator in Eqs.~\eqref{eqn:reg1}~-~\eqref{eqn:reg2}. The dashed line corresponds to the phase shift with $a_0^{1/2} = 0.18 \,  m_\pi^{-1}$. Bottom: Similarly for the S-wave, isospin $I=3/2$ channel. In this case, the dashed line corresponds to the phase shift with $a_0^{3/2} = 0.045 \,  m_\pi^{-1}$. 
 }
 \label{fig:ps}
\end{figure}

\noindent Using the parameters obtained by Ishida \textit{et al.} \cite{Ishida:1997wn} (summarized in Table~\ref{tab:table1}), Eqs.~\eqref{eqn:ps_para1}~-~\eqref{eqn:ps_para3} provide a good description of the experimental data up to $1.6 \, {\rm GeV}$, as shown in Fig.~\ref{fig:ps}. 

\begin{table}
\begin{center}
\begin{tabular}{|c|c|c|c|c|}
      \hline
      $\alpha_\kappa ({\rm GeV^2})$ & $ M_\kappa ({\rm GeV})$ & $ \alpha_{K_0^*} ({\rm GeV^2})$ & $ M_{K_0^*} ({\rm GeV})$ & $ r^{I=1/2}_c ({\rm GeV^{-1}})  $ \\
       \hline
       $ 3.0098   $ &  $ 0.905   $ & $ 1.437  $ & $ 1.41 $ & $ 3.57  $ \\
       \hline
\end{tabular}
\end{center}

\begin{tabular}{|c|}
      \hline
      $ r^{I=3/2}_c ({\rm GeV^{-1}})  $ \\
       \hline
      $ 0.81  $ \\
       \hline
\end{tabular}

\caption{Parameters used to model the $K$$\pi$ scattering phase shifts in the S-wave, isospin $I=1/2$ and $I=3/2$ channels \cite{Ishida:1997wn}.}
\label{tab:table1}
\end{table}

\begin{table}
\begin{center}
\begin{tabular}{|c|c|c|}
      \hline
      &$ a^{I=1/2}_0 \, m_{\pi} $ & $ a^{I=3/2}_0 \, m_{\pi} $\\
       \hline
     Ishida \textit{et al.} \cite{Ishida:1997wn} & $ 0.393  $ & $ -0.112 $\\
       \hline
       B\"uttiker \textit{et al.} \cite{Buettiker:2003pp} & $ 0.224(22) $ & $ -0.045(8) $\\
       \hline
\end{tabular}
\end{center}
\caption{The S-wave $K$$\pi$ scattering lengths obtained with the model \cite{Ishida:1997wn} confronted with an empirical value.}
\label{tab:table2}
\end{table}

An additional constraint on the fit of ${\delta_0^{{1}/{2}}}$ comes from the scattering length $a_0^{1/2}$ \cite{Bugg:2003kj}, which is related to the phase shift near the threshold by

\begin{align}
	\label{eqn:scat_length}
	{\delta_0^{{1}/{2}}}(\sqrt{s} \simeq m_{th}) := a_0^{1/2} P_{CM}(s) + \mathcal{O}(P_{CM}^2). 
\end{align}

\noindent However, the threshold behavior is not uniquely determined by the data. Hence, additional input is needed to obtain an accurate description of the scattering length.

The value for the $I=1/2$ S-wave $K$$\pi$ scattering length obtained in the model is high compared to that obtained in a dispersive analysis of $K$$\pi$ scattering~\cite{Buettiker:2003pp} (see Table II). Low-energy theorems based on the current algebra and the partially conserved axial-vector current (PCAC) predict a lower value of $a_0^{1/2} \approx 0.14 \, m_\pi^{-1}$ \cite{ Weinberg:1966kf, Griffith:1969ph, Bernard:1990kw}. At next-to-leading order in chiral perturbation theory~\cite{Bernard:1990kw}, the scattering length is $\approx 0.18 m_\pi^{-1}$, while agreement with the dispersive approach can be obtained at NNLO~\cite{Bijnens:2004bu}. The extraction of the $K$$\pi$ scattering length from lattice QCD is at present not conclusive. So far, such calculations were done with large pion masses, resulting in large values of the scattering lengths~\cite{Lang:2012sv, Fu:2011wc}. The extrapolation to physical value of the pion mass is delicate. Using chiral perturbation theory at next-to-leading order for the extrapolation, Fu~\cite{Fu:2011wc} finds a somewhat low value, $a_0^{1/2}\simeq 0.18 m_\pi^{-1}$.  

To cover the range of uncertainty in $a_0^{1/2}$, we introduce a regulator 

\begin{align}
	\label{eqn:reg1}
	\mathcal{F}(s) = 1 - \frac{f_0}{1+s/\Lambda^2}
\end{align}

\noindent in the $\kappa$-contribution to the phase shift $\delta_\kappa$, such that

\begin{align}
	\label{eqn:reg2}
	\delta_{\kappa}(s) &= \tan^{-1}\left[\mathcal{F}(s) \frac{-\sqrt{s}\, \gamma(s)}{s-M_0^2}\right]. 
\end{align}

\noindent Numerically we use $\Lambda = 0.381 \, {\rm GeV}$ and vary $f_0$ to obtain a scattering length between $0.18 \, m_\pi^{-1}$ for $f_0=1$ and $ 0.4 \, m_\pi^{-1}$ for $f_0=0$.
	
Before we end the discussion of phase shift, we comment on two important features. First, the shape of ${\delta_0^{{1}/{2}}}(s)$ differs qualitatively from that of a narrow resonance. In the limit of vanishing width, as usually assumed in the HRG model, the phase shift would become a step function which reaches the value of $180^o$ at a mass of $\sqrt{s}\approx 0.682 \, {\rm GeV}$. A comparison with Fig. \ref{fig:ps} clearly shows that the $\kappa$ meson cannot be treated as a narrow resonance. Second, the behavior of the phase shift at threshold is determined by the orbital angular momentum. In an S-wave, the derivative of the phase shift with respect to $s$

\begin{align}
	\frac{d}{ds} \delta^{1/2}_0 \approx a_0^{1/2} P_{CM}^\prime(s \rightarrow m_{th}^2),
\end{align}

\noindent diverges at threshold, due to the fact that $P_{CM}^\prime = d P_{CM}/d s$ diverges at $s = m_{th}^2$ (see Eq.~\eqref{eqn:width_para}). As we discuss in the next section, this fact determines the behavior of the effective spectral weight derived from the phase shifts.

\subsection{S-matrix approach and thermodynamics}

\begin{figure*}[ht]
	\centering
 \includegraphics[width=0.49\textwidth]{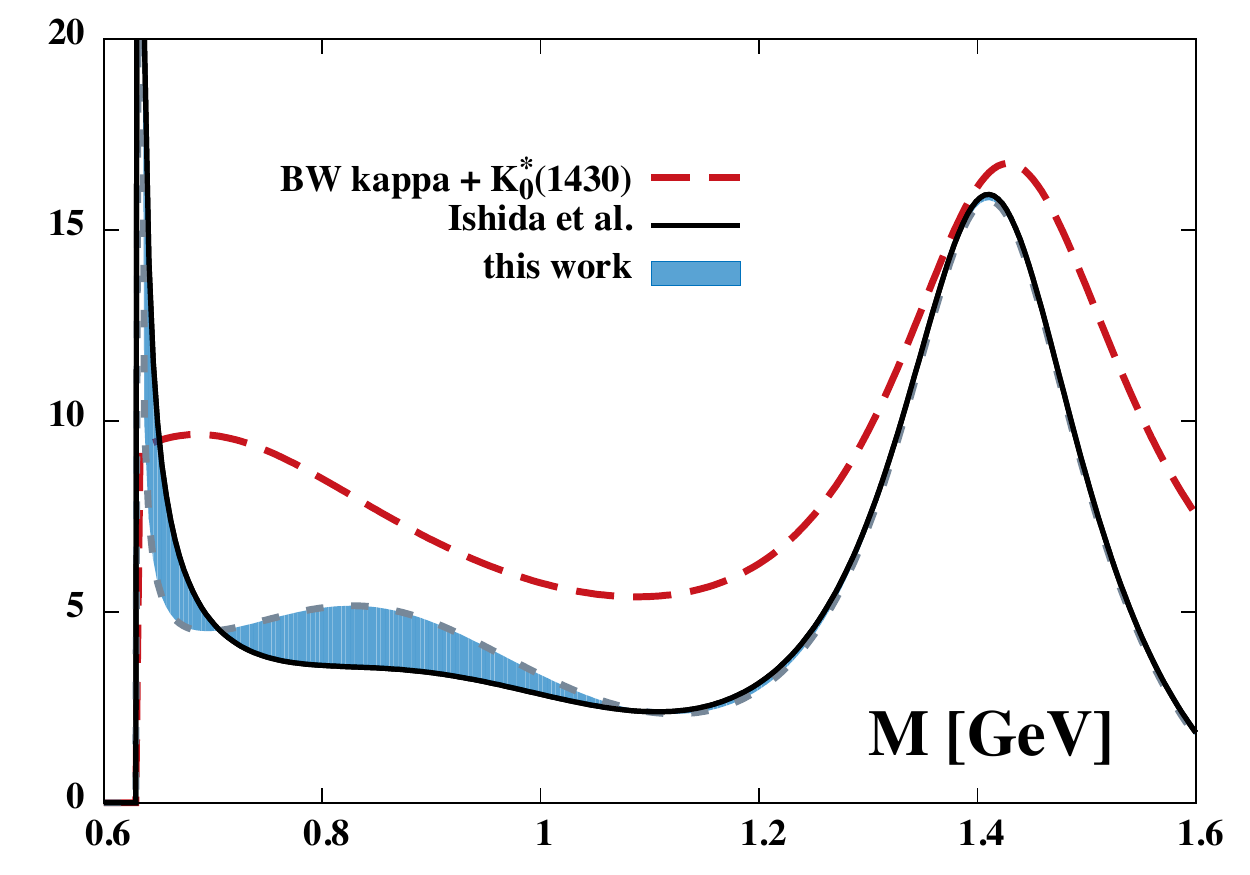}
 \includegraphics[width=0.49\textwidth]{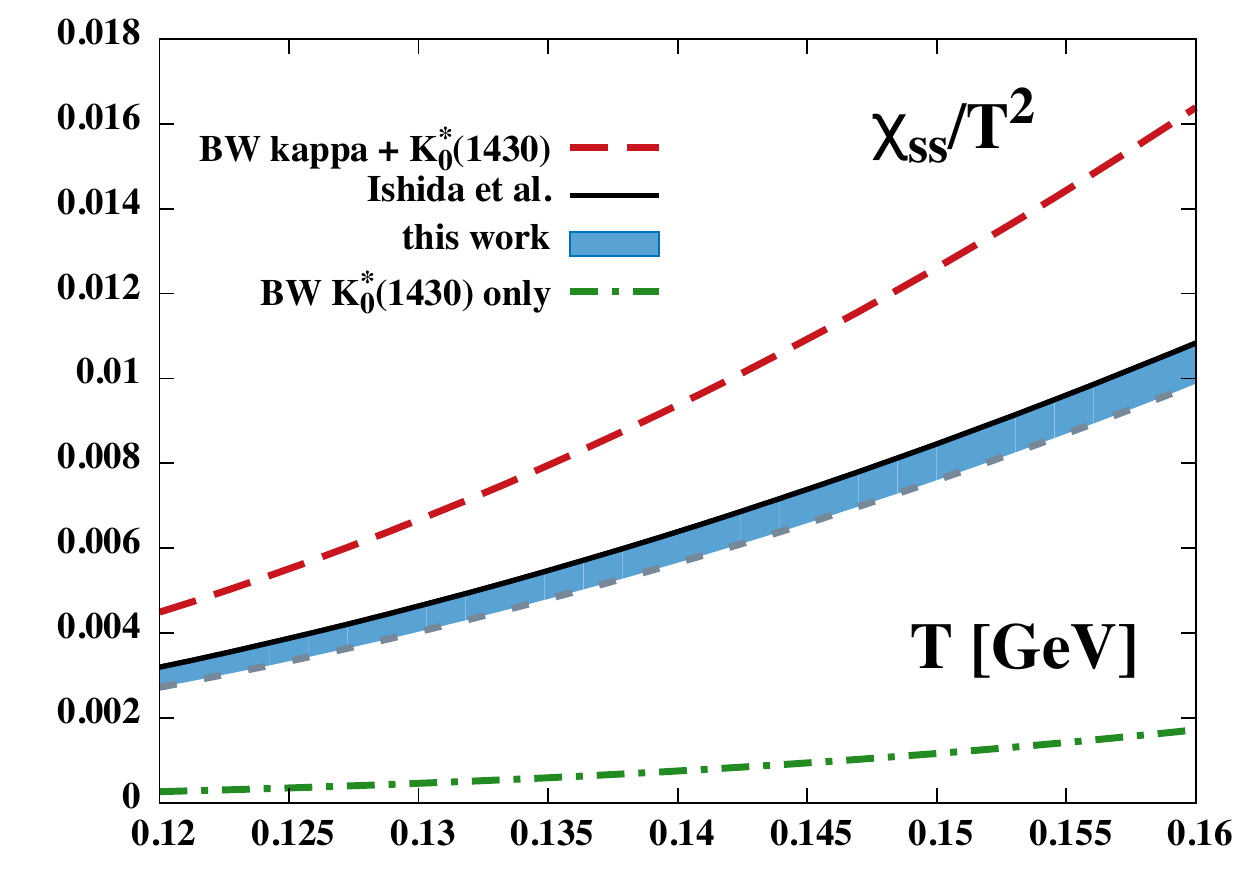}
 \caption{(color online). Left: Comparison of the weight function $\mathcal{B}(M)$ (solid line) and the double Breit-Wigner spectral function for $\kappa$ and $K^*_0(1430)$ (dashed line), both in units of ${\rm GeV}^{-1}$. The band corresponds to different values of the scattering length, bounded by the solid and dashed lines, which corresponds to the scattering length of $a_0^{1/2} = 0.4 \, m_\pi^{-1}$ and $0.18 \,  m_\pi^{-1}$ respectively. Right: The interaction contributions to the strangeness susceptibility in the S-wave $I=1/2$ channel, obtained using different spectral weights.
 }
 \label{fig:k_spec}
\end{figure*}

Given the parametrization of the $K$$\pi$ S-wave phase shifts presented above, we are now ready to formulate the thermodynamics. The tool of choice is the S-matrix approach~\cite{Dashen:1969ep}, which provides a systematic way to account for interactions in a many-body system in thermal equilibrium. The leading order correction, which is determined by the two-body scattering phase shift, is equivalent to the second virial coefficient~\cite{Beth:1937zz}. We apply this formalism to compute the interaction contribution of $K$$\pi$ scattering in the $\kappa$ channel to the thermodynamics of strongly interacting matter in the hadronic phase.

In this approach, the thermodynamic potential $\Omega$ of an interacting system of pions, kaons and resonances is, to leading order, given by the sum:

\begin{align}
	\label{eqn:omega}
	\Omega = \Omega_\pi + \Omega_K + \Omega_{int}. 
\end{align}

\noindent The first two terms are the ideal gas expressions for pions and kaons:

\begin{align}
	\Omega_{\pi} &= 3 T V \int \frac{d^3 p}{(2 \pi)^3} \, \left\{\ln[1 - e^{-\beta \sqrt{p^2+m_\pi^2}}]\right\} \\
	\Omega_{K} &= 2 T V \int \frac{d^3 p}{(2 \pi)^3} \, \left\{\ln[1 - e^{-\beta (\sqrt{p^2+m_K^2}+\mu_S)}]\right.\nonumber\\
	           &+\left.\ln[1 - e^{-\beta (\sqrt{p^2+m_K^2}-\mu_S)}]\right\}, 
	\label{eqn:free_gas}
\end{align}

\noindent where $\mu_S$ is the strangeness chemical potential and the two terms in $\Omega_K$ are due to kaons and antikaons, respectively. Finally, the last term in Eq.~\eqref{eqn:omega} accounts for $K$$\pi$ interactions. In the HRG approach, $\Omega_{\rm int}$ is given by the sum of all relevant resonances treated as an ideal gas of stable particles:

\begin{align}
	\Omega^{\rm HRG}_{\rm int} &= \sum_{\rm res.} 2 T V \int \frac{d^3 p}{(2 \pi)^3} \, \left\{\ln[1 - e^{-\beta (\sqrt{p^2+m_i^2}+\mu_S)}]\right. \nonumber \ \\&+\left.\ln[1 - e^{-\beta (\sqrt{p^2+m_i^2}-\mu_S)}]\right\}
\end{align}

\noindent The degeneracy factor accounts for the two possible isospin states in the $I(J^P)=\frac{1}{2}(0^+)$ channel. The thermodynamic pressure is computed by

\begin{align}
	P = -\frac{ \Omega }{V}.
\end{align}

\noindent Another key quantity of interest is the strangeness susceptibility, which is obtained by taking derivatives of the thermodynamic pressure with respect to the strangeness chemical potential

\begin{align}
	{\chi}_{SS} = \frac{\partial^2 P}{\partial {\mu}_S \partial {{\mu}}_S} \biggr \vert_{\mu_S = 0}.
\end{align}

\noindent The results for the thermodynamic observables in HRG are shown in Fig.~\ref{fig:hrg}.

In the S-matrix approach, the interaction contribution to the thermodynamic potential involves an integral over the invariant mass $M=\sqrt{s}$:

\begin{align}
\label{eqn:beth_potential}
\Omega^{\rm B}_{\rm int} \approx 2 T V &\int_{m_{th}}^\infty \frac{dM}{2 \pi} \int \frac{d^3 p}{(2 \pi)^3} \, \mathcal{B}(M)\nonumber \\
&\times \left\{\ln[1 - e^{-\beta (\sqrt{p^2+M^2}+\mu_S)}]\right.\\
&+\left.\ln[1 - e^{-\beta (\sqrt{p^2+M^2}-\mu_S)}]\right\}. \nonumber
\end{align}
 
\noindent with the effective weight function~\cite{Dashen:1969ep,Weinhold:1997ig}

\begin{align}
	\label{eqn:beth}
	\mathcal{B}(M) &= 2 \frac{d}{d M} \delta(M),
\end{align}

\noindent which satisfies the normalization condition

\begin{align}
\int_{m_{th}}^\infty \frac{dM}{2 \pi}  \, \mathcal{B}(M) = 1,
\end{align}

\noindent provided the phase shift has the property $\delta(\infty) \rightarrow \pi$. We note that the weight function $\mathcal{B}$ is in principle defined in any channel, irrespective of the existence of a corresponding resonance. Moreover, even for a well defined resonance, the weight function differs from the corresponding spectral function~\cite{Weinhold:1997ig}. The two functions are identical only in the limit, where the width of the resonance vanishes.

Using the thermodynamic potential (\ref{eqn:beth_potential}), one can compute the interaction contribution to thermodynamic observables. Thus, e.g. the interaction contribution to the thermodynamic pressure is given by

\begin{align}
\label{eqn:pressure}
{\hat{P}}_{int} &= \int_{m_{th}}^\infty \frac{dM}{2 \pi} \mathcal{B}(M) \hat{P}_T(M),
\end{align}

\noindent where

\begin{align}
\label{eqn:pressure_2}
\hat{P}_T(M) &= -2 \int \frac{d^3 \hat{p}}{(2 \pi)^3} \left\{\ln[1 - e^{-\sqrt{\hat{p}^2+\hat{M}^2}-\hat{\mu}_S}]\right.\\ \nonumber
&+\left.\ln[1 - e^{-\sqrt{\hat{p}^2+\hat{M}^2}+\hat{\mu}_S}]\right\} 
\end{align}

\noindent with $\hat{P} = P/T^4$, $\hat{p} = p/T$, $\hat{M}=M/T$ and $\hat{\mu}_S = \mu_S/T$. The interaction effects on the strangeness susceptibility will be discussed in the following section.

\begin{figure*}[ht]
	\centering
 \includegraphics[width=0.49\textwidth]{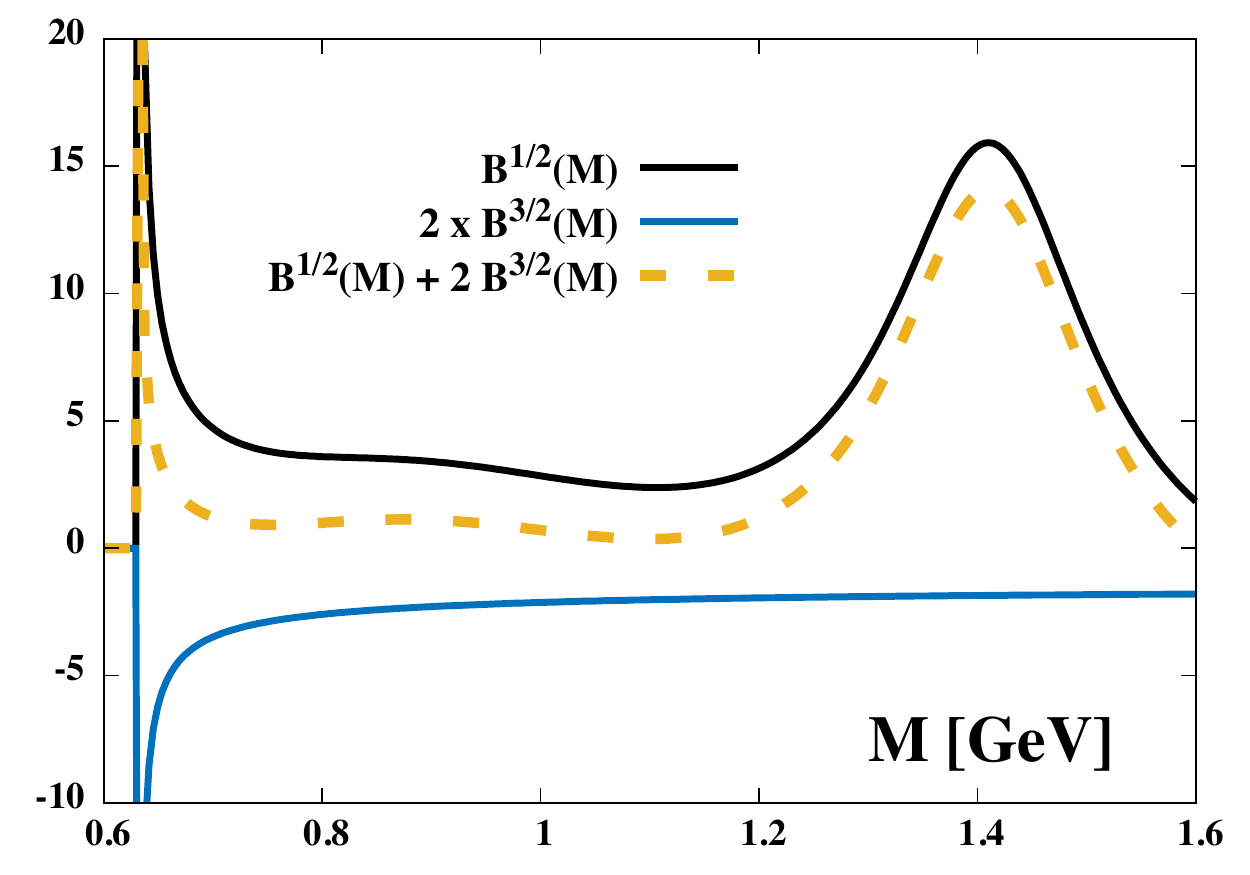}
 \includegraphics[width=0.49\textwidth]{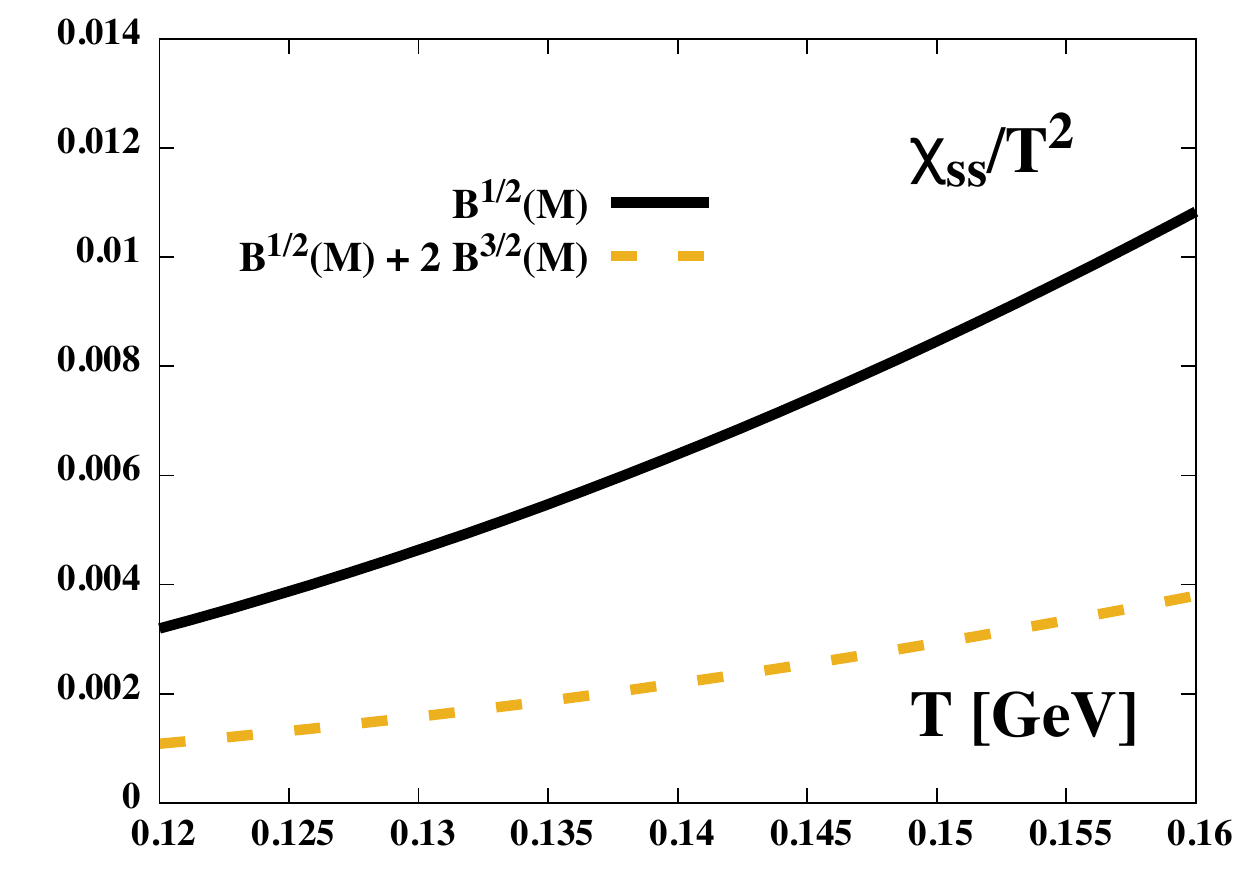}
 \caption{(color online). Left: Comparison of the S-wave weight functions $\mathcal{B}^I(M)$ (${\rm GeV}^{-1}$) in different isospin channels and their weighed sum. 
	 Right: The interaction contributions to the strangeness susceptibility (normalized to $T^2$) from the $I=1/2$ and $I=3/2$ S-wave channels.
 }
 \label{fig:iso32}
\end{figure*}

\section{Influence of the $\kappa$ channel}

The parametrization of the phase shifts and weight function presented above allow for an assessment of various approximate  descriptions of the interaction effects in the $K$$\pi$ channel. For example, a standard Breit-Wigner resonance with an energy-independent width $\gamma(s) \rightarrow\gamma_{\rm BW}$ is easily accommodated by neglecting the $s$-dependence of the numerator $ \sqrt{s} \gamma(s) $ in the phase shift formula Eq.~\eqref{eqn:ps_para1} \cite{Agashe:2014kda}, thus

\begin{align}
	\label{eqn:beth_to_BW}
	\mathcal{B}(M) &= 2 \frac{d}{d M} \delta(M)  \nonumber \\ 
		       &\rightarrow 2 M \frac{2 M \gamma_{\rm BW}}{(M^2-M_0^2)^2 + M^2 \gamma_{\rm BW}^2}.
\end{align}

The weight functions in different approximation schemes are shown in Fig.~\ref{fig:k_spec} (left panel). The validity of the weight function is limited to energies below $M=1.6 \,{\rm GeV}$, the highest energy included in the fit of the phase shifts. For the computation of thermodynamic observables, e.g. pressure in Eqs.~\eqref{eqn:pressure}~-~\eqref{eqn:pressure_2}, the integral over $M$ converges well below this energy for temperatures up to $0.16 \,{\rm GeV}$, owing to the suppression by the Boltzmann factor.

One characteristic feature of the weight function $\mathcal{B}$ for an S-wave channel is that it diverges at the threshold, as seen in the left panel of Fig.~\ref{fig:k_spec}. This singularity is, however, integrable, and its sign as well as its strength are directly related to the scattering length in the corresponding partial wave. A lower value of the scattering length tends to reduce the strength of the weight function near the threshold, shown as the blue band of Fig.~\ref{fig:k_spec}. The standard Breit-Wigner form, on the other hand, does not exhibit such a divergence.

In addition, in Fig.~\ref{fig:k_spec} (left panel), we observe a secondary peak appeared near $ 1.4 \, {\rm GeV}$. This clearly corresponds to the $K^*_0(1430)$ resonance.

In the right panel of Fig.~\ref{fig:k_spec} we show the dependence of $\chi_{SS}$ on the weight function. We observe that the S-matrix approach yields a result that lies between those obtained within the standard Breit-Wigner, for $K^*_0(1430)$ with and without $\kappa$. Moreover, the dependence of $\chi_{SS}$ on the scattering length is displayed. We see that a larger positive scattering length provides more support in the low mass region and hence gives an enhanced contribution to the susceptibility. 

We now discuss the origin of the suppression of thermodynamic observables when treating the interaction based on $\kappa$-channel phase shift. Previous studies have stressed the importance of using $\mathcal{B}$ instead of the standard spectral function \cite{Weinhold:1997ig}. For the case of $\Delta$ resonance, where P-wave scattering is involved, the $\mathcal{B}$ function tends to enhance the low mass contribution to the thermodynamics and results in an overall increase in the observables beyond those treated by the standard Breit-Wigner approach. In the current study involving an S-wave scattering, the enhancement effect near the threshold is, however, compensated by the relatively slow increase in the phase shift below $1.3 \, {\rm GeV}$. Unlike a typical resonance, the phase shift in the $\kappa$ channel does not reach $180^o$ before $K^*_0(1430)$ emerges. The slow rise of the phase shift in the low mass region limits the strength of weight function $\mathcal{B}$. Consequently, the thermodynamic observables calculated in the S-matrix approach for $\kappa$ becomes strongly suppressed, making its contribution too small to remove the disparity between HRG and lattice results (see Fig.~\ref{fig:hrg}).

Thus, our calculation shows that the contribution of $K\pi$ interactions to the strange susceptibility is substantially lower in the current consistent treatment compared to the HRG description. As a result, the inclusion of the $\kappa$ channel does not resolve the issue of the missing strength in the strange sector. Clearly, a careful analysis of the interaction strength in other strangeness carrying channels is called for.

\section{The effect of $I = 3/2$ $K$$\pi$ scattering}

For completeness, we also assess the contribution of the isospin $I = 3/2$ S-wave channel to the strangeness fluctuations. As pointed out in Ref. \cite{Broniowski}, the inclusion of this channel partly cancels the effect of the isospin $I=1/2$ channel. This is expected since the phase shift in the $I = 3/2$ channel, and consequently the weight function $\mathcal{B}^{I=3/2}(M)$, are negative, corresponding to a repulsive interaction. 

Given that this channel involves only non-resonant $K$$\pi$ scattering, we employ the repulsive core expression for the phase shift

\begin{align}
	\delta^{3/2}_{0}(M) = - r^{I=3/2}_c P_{CM}(M).
	\label{eqn:ps32}
\end{align}

\noindent The fit parameter $r^{I=3/2}_c = 0.112 \, m_\pi^{-1}$, suggested by Ishida \textit{et al.} \cite{Ishida:1997wn}, yields a high value for the scattering length compared to that obtained in the dispersive analysis of \cite{Buettiker:2003pp} (see Table II). To cover the range of uncertainties in the scattering lengths, we employ the regulator introduced in Eq.~\eqref{eqn:reg1}. We use $\Lambda = 0.381 \, {\rm GeV}$ and dial $f_0$ to obtain a scattering length between $0.045 \, m_\pi^{-1}$ for $f_0=2.28$ and $ 0.112 \, m_\pi^{-1}$ for $f_0=0$.

The resulting phase shift $\delta^{3/2}_0$ is shown in Fig.~\ref{fig:ps} (lower panel). It is evident that Eq.~\eqref{eqn:ps32} cannot capture the full features of the phase shift, and a more refined approach (like e.g. \cite{Barnes:1992qa}) can achieve a more satisfactory description of the non-resonant scattering. Nevertheless, the parametrization~\eqref{eqn:ps32} is sufficient for our current discussion.

The weight function $\mathcal{B}$, in the thermodynamic potential discussed in Eqs.~\eqref{eqn:beth_potential}~-~\eqref{eqn:pressure_2}, is then modified as follows:

\begin{align}
	\mathcal{B} &= \mathcal{B}^{I=1/2} + 2 \,  \mathcal{B}^{I=3/2}  \nonumber \\
	\mathcal{B}^{I}  &= 2 \frac{d}{d M}\mathcal{\delta}^{I}(M),
\end{align}

\noindent where $\mathcal{B}^I(M)$ is the weight function in isospin channel $I$ and the factor of $2$ in front of $\mathcal{B}^{I=3/2}$ accounts for the relative isospin degeneracy factor of the $I=3/2$ and $I=1/2$ channels. 

The corresponding contributions to the weight function and the strangeness susceptibility are shown in Fig.~\ref{fig:iso32}. 
The partial cancellation between the two isospin channels is evident. As a result, the enhancement of strangeness fluctuations due to S-wave $K\pi$ scattering is reduced by $70 \%$. This effect, which is not  accounted for in the HRG model, lends further support to our conclusion that a consistent treatment of low-mass resonances requires a careful analysis, including also non-resonant interactions. A natural framework for such studies is offered by the S-matrix approach employed in this paper.

\section{Conclusion}

This study set out to explore possible sources of missing strength in the strangeness susceptibility, suggested by lattice results. The $K^*_0(800)$ resonance, a.k.a. $\kappa$, which is not an established resonance in the PDG compilation, appears to be a promising candidate. Indeed, within the treatment of the hadron resonance gas (HRG) model, we found that this single state alone accounts for the missing contribution in the strange susceptibility.

However, owing to the large width of the $\kappa$ meson and the significant non-resonant background, the HRG model does not provide an accurate description of the interaction contributions. In fact, a consistent treatment of all $K\pi$ S-wave interactions within the S-matrix approach shows that 
a simplified (HRG) treatment of the interactions in these channels, using a Breit-Wigner spectral function for each resonance and ignoring the non-resonant background, systematically overestimates the contribution to strangeness fluctuations.
	
In summary, the $K\pi$ S-wave interactions provide only a part of the missing contribution to the strangeness susceptibility, indicated by recent lattice QCD results. Whether the remaining discrepancy can be resolved by a consistent treatment of other strangeness carrying channels will be explored in future studies.

\acknowledgments

We acknowledge the stimulating discussions with A. Andronic, D. Blaschke, P. Braun-Munzinger,  F. Karsch, J. Stachel and L. Turko. B. F. is supported in part by the Extreme Matter Institute EMMI. C. S. acknowledges partial support of the Hessian LOEWE initiative through the Helmholtz International Center for FAIR (HIC for FAIR). This work was partly supported by the Polish National Science Center (NCN), under Maestro grant DEC-2013/10/A/ST2/00106.

\end{document}